\documentclass[aps,pra,superscriptaddress,amsmath,amssymb,showpacs]{revtex4-2}
\usepackage[english]{babel}
\usepackage[cp866]{inputenc}
\usepackage{amssymb,bm}
\usepackage{graphicx}
\usepackage{amsmath}
\usepackage{color}
\usepackage[colorlinks=true,citecolor=blue,urlcolor=blue]{hyperref}
\begin{document}

	\title{Coulomb effects in the decays   $\Upsilon (4S) \rightarrow B\bar B$ }

	\author{A.I. Milstein}
	\email{A.I.Milstein@inp.nsk.su}
	\author{S.G. Salnikov}
	\email{S.G.Salnikov@inp.nsk.su}
	\affiliation{Budker Institute of Nuclear Physics of SB RAS, 630090 Novosibirsk, Russia}
	\affiliation{Novosibirsk State University, 630090 Novosibirsk, Russia}
	\date{\today}

	\begin{abstract}
	A simple exactly solvable model is proposed for describing the decays  $\Upsilon (4S) \rightarrow B^0\bar B^0$  and  $\Upsilon (4S) \rightarrow B^+B^-$. Our predictions agree with available experimental data. Using this model, we analyze the  Coulomb effects in the spectra of these decays. It is shown that the frequently used assumption of factorization of Coulomb effects is not fulfilled. The Coulomb interaction leads to the difference in the positions and  heights  of the peaks corresponding to  the charged and neutral modes. As a result, the ratio of  probability of $\Upsilon (4S)\rightarrow B^+B^-$ decay and $\Upsilon (4S) \rightarrow B^0\bar B^0$ decay is a nontrivial function of energy. 
	\end{abstract}
	\maketitle
	
\section{Introduction}
The study of resonances of mass $ M_R $, which slightly exceeds the particle-antiparticle pair production threshold $M_{th} $, is a very important task. Such a study makes it possible to investigate in detail the effects of strong interaction in the region where perturbation theory is not applicable. Of particular interest is the case when $ M_R-M_{th} $ is of the order of the resonance width $\Gamma_R $, since in this case the shape of the resonance curve becomes very nontrivial. In addition, the low relative velocity of a pair of charged particles makes the influence of Coulomb effects very important. In particular, these conditions correspond to the decays of $\Upsilon (4S) \rightarrow B^0\bar B^0$ and  $\Upsilon (4S) \rightarrow B^+B^-$. 
 
 For resonance $\Upsilon (4S)$, we have $M_R=10579.4\pm 1.4\,\mbox{MeV}$, $M_R-M_{th}= 20.1\pm 1.4\,\mbox{MeV}$ for neutral mesons and $ 20.7 \pm 1.4 \, \mbox {MeV} $ for charged mesons, $\Gamma_R= 20.5\pm 2.5 \,\mbox{MeV}$, and the sum of the probabilities $W_c$ and $W_n$ of the decays $\Upsilon (4S) \rightarrow B^+B^-$ and  $\Upsilon (4S) \rightarrow B^0\bar B^0$, respectively, is almost 100\%. Note that the mass difference $M_{B^0}-M_{B^+}\approx0.3 \,\mbox{MeV}\ll\Gamma_R$, and in the first approximation this difference can be neglected. 

Currently, there are a number of experimental works devoted to the decays of the $\Upsilon (4S)$ meson \mbox{\cite{CLEO2001, CLEO2002, BABAR2002, Belle2003, BABAR2004, BABAR2005branching, BABAR2005width, BABAR2009, Dong2020, Belle2021}}. The parameter $t=\pi\alpha/v\approx 0.37$, which determines the magnitude of the Coulomb effects, is not small (here $v$ is the velocity of the $ B $ meson, $ \alpha $ is the fine-structure constant, $ \hbar = c = 1 $). Therefore, it would be possible to estimate the magnitude of the Coulomb effects using the Sommerfeld-Gamow-Sakharov factor, $W_c/W_n\sim t/[1-\exp(-t)]\approx 1.2$. However, the value of this ratio given in the PDG tables \cite{PDG2020} is much smaller, $W_c/W_n= 1.058 \pm 0.024\,$. Various theoretical approaches have been proposed to estimate Coulomb effects in $\Upsilon (4S) \rightarrow B\bar B$ decays with completely different qualitative predictions of the magnitude of these effects~\cite{Marciano1990, Lepage1990, Eichten1990, Kaizer2003, Voloshin2003, Voloshin2005}.

The appearance of experimental data for the resonance shape of $\Upsilon (4S)$ \cite{BABAR2009, Dong2020} allows one to make a progress in understanding Coulomb effects. Unfortunately, these experimental data contain information only on the shape of the spectral line for the sum $W_c+W_n$, and the ratio $W_c/W_n$ is measured only at the energy corresponding to the maximum of this sum. However, there is a question  whether the positions of peaks in the spectrum of charged and neutral $B$ mesons coincides, and how the Coulomb interaction affects the shape and position of these peaks. 

In our paper, we have suggested a simple exactly solvable model that describes the available experimental data  and allows us to answer, at least qualitatively, the questions mentioned above. Of course, at present it is not known much about the exact interaction Hamiltonian of $B$-mesons. However, for our purposes, this is  not a big problem. This is similar to the situation with the phenomenological description of the quarkonium spectra, where completely different analytical representations of the potentials are used, and all of them describe both the quarkonium spectra and the wave functions well enough. Therefore, we expect our predictions to clarify the physics of near-threshold resonances and explain the meaning of  observed effects.

\section{Theoretical approach}

Technically, the method for solving the problem under discussion is similar to that developed for calculating the cross section of $e^+e^-$ annihilation into  proton-antiproton  and  neutron-antineutron pairs near the pair production thresholds \cite{DMS2014, DMS2016, MS2018}. The main difference of these two problems, in addition to different spins, angular momenta and a significant mass difference of the proton and neutron, is high probability of meson production near the nucleon-antinucleon pair production threshold. Therefore, 
 in $\Upsilon (4S) \rightarrow B\bar B$ decay we can use the usual potential instead of
an optical potential, as in the case of nucleon-antinucleon pair production.

In our problem $B\bar B$ pair is produced in a state with an orbital angular momentum $l=1$. At small distances, a pair of $b\bar b$ quarks is produced in a state with zero isospin, which, as a result of hadronization, transforms into a superposition of interacting $B^+B^-$ and $B^0\bar B^0$ mesons. Due to the electromagnetic interaction of charged mesons, a state with isospin zero is admixed with a state of isospin one. As a result, the probabilities of charged and neutral meson pair production are different. Of course, the difference between the masses of $B^+$ and $B^0$  also leads to the isospin violation, but this difference is very small ($\sim 0.3\, \mbox{MeV}$), and can be ignored in the first approximation. The Coulomb interaction of $b$ quarks at small distances also leads to a very small effect. Hence, the effect of isospin symmetry violation  is mainly related to the Coulomb interaction of mesons in the final state.

Following \cite{MS2018}, consider the radial wave function $\Psi^T(r)=\left(U^{(c)}(r),\,U^{(n)}(r)\right)$ of  $B\bar B$ pair, where $U^{(c)}(r)$ corresponds to a pair of charged $B$-mesons, and $U^{(n)}(r)$ corresponds to a pair of neutral $B$-mesons, $T$~means transpose. It is convenient to pass
from the wave function $\Psi^T (r)$ to the wave function $\psi^T(r)=kr\,\Psi^T(r)=\left(u^{(c)}(r),\,u^{(n)}(r)\right)$, where $k=\sqrt{M_BE}$, $M_B$ is the mass of the meson, and $E$ is the energy of $B\bar B$ pair, counted from $M_{th}=2M_B$. The function $\psi(r)$  satisfies the equation
\begin{align}\label{we}
& \left[-\dfrac{1}{M_B}\,\dfrac{\partial^2}{\partial\,r^2}+
\dfrac{2}{M_Br^2}+V(r)+ V_{ex}(r)\,\begin{pmatrix}0 &1\\
1 &0\end{pmatrix}-\dfrac{\alpha}{r}\,\begin{pmatrix}1 &0\\
0&0\end{pmatrix}-E \right]\psi(r)=0\,,\nonumber\\
&V(r)=\dfrac{1}{2}[{\cal V}_1(r)+{\cal V}_0(r)]\,,\quad V_{ex}(r)=\dfrac{1}{2}[{\cal V}_1(r)-{\cal V}_0(r)]\,,
\end{align}
where  ${\cal V}_1(r)$ and ${\cal V}_0(r)$ are the potentials of  meson interaction  in the states with isospin one and zero, respectively. The  potential $ V_{ex}(r)$ leads to the transitions  $B^+B^-\leftrightarrow B^0\bar B^0$. 
 
 It is necessary to find two solutions $\psi_i(r)$ of \eqref{we} with asymptotics at large distances
\begin{align}
& \psi_{1}^{T}(r)=\frac{1}{2i}\left(S_{11}\chi_{c}^{+}-\chi_{c}^{-},\,S_{12}\chi_{n}^{+}\right),\nonumber \\
& \psi_{2}^{T}(r)=\frac{1}{2i}\left(S_{21}\chi_{c}^{+},\,S_{22}\chi_{n}^{+}-\chi_{n}^{-}\right),
\end{align}
Here $S_{ij}$ are some functions of energy and 
\begin{align}
& \chi_{c}^{\pm}=\exp\left\{\pm i\left[kr- \pi/2+\eta_k\ln(2kr)+\sigma_k\right]\right\},\nonumber \\
& \chi_{n}^{\pm}=\exp\left[\vphantom{\bigl(\bigr)}\pm i\left(kr- \pi/2\right)\right],\nonumber \\
& \sigma_k=\frac{i}{2}\ln\frac{\Gamma\left(2+i\eta_k\right)}{\Gamma\left(2-i\eta_k\right)}\,,\qquad
\eta_k=\frac{M_B\alpha}{2k}\,,
\end{align}
where  $\Gamma(x)$ is the Euler $\Gamma$ function.

The probabilities $W_c$ and $W_n$ of decays $\Upsilon(4s)\rightarrow B^+B^-$ and $\Upsilon(4s)\rightarrow B^0\bar B^0$, respectively,  are
 \begin{align}\label{3W}
&W_c=N\,k\,\left|\dfrac{\partial}{\partial r}\,U_{1}^{(c)}(0)-\dfrac{\partial}{\partial r}U_{1}^{(n)}(0)\right|^2\,,\nonumber\\
&W_n=N\,k\,\left|\dfrac{\partial}{\partial r}U_{2 }^{(c)}(0)-\dfrac{\partial}{\partial r}U_{2 }^{(n)}(0)\right|^2\,,
\end{align}
where $N$ is some constant. As a model potential, we choose $V(r)=-V_0\,\theta(a-r)$ and 
$V_{ex}(r)=g\,\delta(r-a)$, where $\theta(x)$ is the Heaviside function, $\delta(x)$ is the Dirac $\delta$-function, $ V_0 $, $ g $ and $a$ are some parameters. It turned out that such a simple model is sufficient to describe the available experimental data well enough. Using this potential model, it is easy to obtain an analytical solution, which simplifies the analysis of the influence of Coulomb effects on the probability of pair production. We are confident that, at least qualitatively, our predictions correspond to the actual experimental situation.

For $r<a$, the solutions are regular at the point $r=0$ and have the form
\begin{align}
&u_{1,2}^{(c)}(r)=A_{1,2}\,{\cal F}(y)\,,\quad u_{1,2}^{(n)}(r)=B_{1,2}\,f(y)\,,\nonumber\\
&{\cal F}(y)=\dfrac{C_q}{3}y^2 \,e^{-iy}\,F(i\eta_q+2,4,\,2iy)\,,\quad
f(y)=\dfrac{\sin y}{y}-\cos y\,,\nonumber\\
&C_q=\sqrt{\dfrac{2\pi\eta_q\,(1+\eta_q^2)}{1-\exp(-2\pi\eta_q)}}\,,\quad
 y=qr\,,\quad q=\sqrt{M_B(E+V_0)}\,.
\end{align}
Here  $F(b,c,z)$  is the confluent hypergeometric function of the first kind, $A_{1,2}$ and $B_{1,2}$ are some coefficients. For $r>a$ the solutions are
\begin{align}
&u_{1}^{(c)}(r)=\frac{1}{2i}\left[S_{11}\,H^{+}(k,x)  -H^{-}(k,x)\right],\quad u_{1}^{(n)}(r)=  \frac{1}{2i}\,S_{12}\,h^{+}(x)\,,\nonumber\\
&u_{2}^{(c)}=\frac{1}{2i}\,S_{21}\,H^{+}(k,x)\,,\quad u_{2}^{(n)}(r)=  \frac{1}{2i}\left[\,S_{22}\,h^{+}(x)-h^{-}(x)\right],\nonumber\\
&H^+(k,x)=4i\exp[ix +i\sigma_k-\pi\eta_k/2]\,x^2\,{\cal U}(2-i\eta_k,4,\,-2ix)\,,\nonumber\\
&H^-(k,x)=-4i\exp[-ix -i\sigma_k-\pi\eta_k/2]\,x^2\,{\cal U}(2+i\eta_k,4,\,2ix)\,,\nonumber\\
&h^+(x)=\left(\dfrac{1}{x}-i\right)\,e^{ix}\,,\quad h^-(x)=\left(\dfrac{1}{x}+i\right)\,e^{-ix}\,,\nonumber\\
& x=kr\,,\quad k=\sqrt{M_BE}\,.
\end{align}
Here ${\cal U}(b,c,z)$ is the confluent hypergeometric function of the second kind. The following relations hold
$${\cal F}(y)=\frac{1}{2i}\left[\,H^{+}(q,y)  -H^{-}(q,y)\right],\quad  f(y)=\frac{1}{2i}\left[\,h^{+}(y)  -h^{-}(y)\right].$$
Using the continuity of the function $\psi(r)$ at the point $r=a$ and the condition
 $$\psi'(a+0)-\psi'(a-0)=M_B\,g\,\,\begin{pmatrix}0 &1\\
1 &0\end{pmatrix}\psi(a)\,,$$
we find the coefficients $A_i$  and $Bi$. As a result, we obtain  the probabilities $W_c$ and $W_n$  
\begin{align}\label{wcnfinal}
&W_c=b\, \dfrac{k\,q^2}{M_B^3}\,\Big|\dfrac{q}{D}\Big\{C_q\,\Big[k\,h^{+}{}'(ka)\,f(qa) -q\, h^+(ka)\,f'(qa)\Big]-M_B\,g\,h^+(ka)\,{\cal F}(qa)\Big\}\Big|^2\,,\nonumber\\
&W_n=b\, \dfrac{k\,q^2}{M_B^3}\,\Big|\dfrac{q}{D}\Big\{\Big[k\,H^{+}{}'(ka)\,{\cal F}(qa) -q\, H^+(ka)\,{\cal F}'(qa)\Big]-C_q\,M_B\,g\,H^+(ka)\,f(qa)\Big\}\Big|^2\,,\nonumber\\
&D=\Big[k\,h^{+}{}'(ka)\,f(qa) -q\, h^+(ka)\,f'(qa)\Big]\Big[k\,H^{+}{}'(ka)\,{\cal F}(qa) -q\, H^+(ka)\,{\cal F}'(qa)\Big]\nonumber\\
&-M_B^2g^2\,h^+(ka)\,H^+(ka)\,f(qa)\,{\cal F}(qa)\,,
\end{align}
where $b$ is some constant and $Z'(x)\equiv \partial Z(x)/\partial x$. It is the ratio $R$ of the cross section $e^+e^-\to B\bar B$ to the Born cross section $e^+e^-\to \mu^+\mu^-$, that is usually presented in the experimental papers. Taking into account slow energy dependence of the cross section of $\mu^+\mu^-$ production in the energy region considered, we choose the constant $b$ to reproduce this ratio~$R$. The expressions \eqref{wcnfinal}  are exact within the model under consideration and are very convenient for analyzing various effects.

\section{Discussion of the results}
It turned out that our predictions are rather sensitive to the value of parameter $g$. However,
comparison with experimental data shows that $g$ is very small. Below we put
$g=0$, so the expressions~\eqref{wcnfinal} become much simpler
\begin{align}\label{wcnfinal0}
&W_c=b\, \dfrac{k\,q^2}{M_B^3}\,\Bigg|\dfrac{q\,C_q}{k\,H^{+}{}'(ka)\,{\cal F}(qa) -q\, H^+(ka)\,{\cal F}'(qa)}\Bigg|^2\,,\nonumber\\
&W_n=b\, \dfrac{k\,q^2}{M_B^3}\,\Bigg|\dfrac{q\,}{k\,h^{+}{}'(ka)\,f(qa) -q\, h^+(ka)\,f'(qa)}\Bigg|^2\,.
\end{align}
Our analysis shows that the observed resonance $\Upsilon(4S)$ is related to a low-energy virtual state in the $p$-wave with the potential $V_0$ being much larger than the energy of a virtual level. Therefore, the potential $ V_0 $ can be chosen in the form
\begin{equation}\label{V0}
V_0=\dfrac{(n\pi)^2}{M_Ba^2} -E_R\,,\quad n=3,4,5,6\dots\,,
\end{equation}
where $E_R$ is a parameter close to the value of the resonance energy, $E_R\approx 22 \,\mbox {MeV}\,$. It only slightly depends on $a$ and is almost independent of $n$.
It turned out that for any $a$ in the interval $2\,\mbox{fm}\leqslant a\leqslant 2.5\,\mbox{fm}$  and for $n\geqslant 3$, the curves for $W_c$ and $W_n$, described by Eqs.~\eqref{wcnfinal0} and \eqref{V0}, have similar shapes (up to the general scale $b$), so that the ratio $W_c/W_n$ conserves. Below we use the values $V_0=269\,\mbox{MeV}$, $a=2.5\,\mbox{fm}$,  $b=23$, $g=0$. These values correspond to $n = 5$.

\begin{figure}
	\begin{minipage}[t]{0.48\columnwidth}
		\centering
		\includegraphics[width=\linewidth]{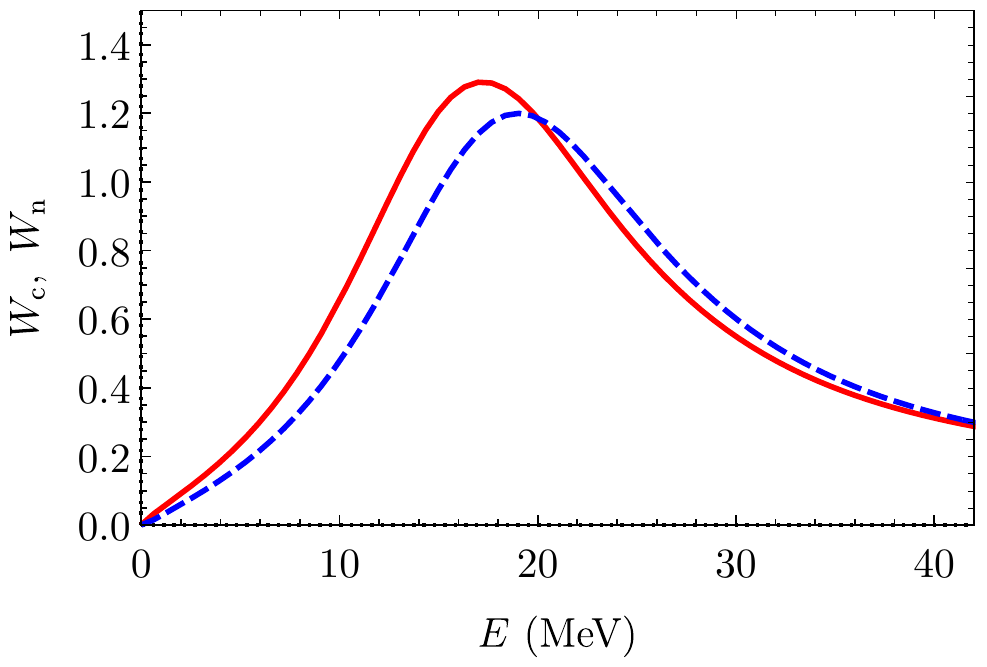}
		\caption{Energy dependence of $W_c$ (solid curve) and  $W_n$ (dotted curve). }
		\label{wcwn}
	\end{minipage}
	\hfill
	\begin{minipage}[t]{0.48\columnwidth}
		\centering
		\includegraphics[width=\linewidth]{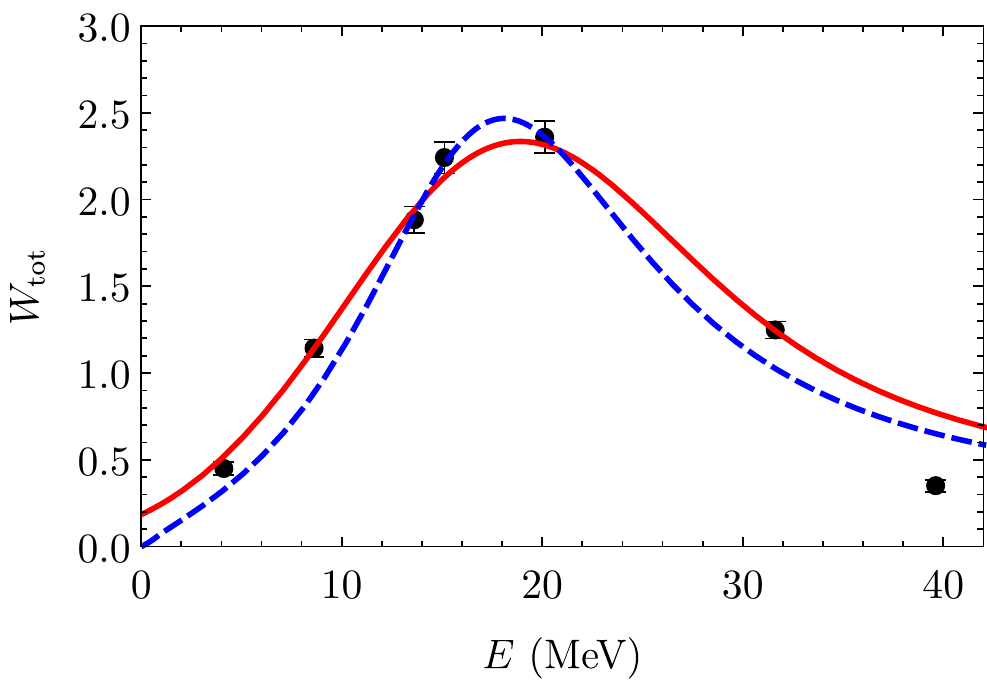}
		\caption{Energy dependence of the probability $W_{tot} = W_c + W_n$ (dashed curve) and $\langle W_{tot}\rangle$ (solid curve). The dots show the experimental data that take into account the radiative corrections and the radiative return \cite{Dong2020} .}
		\label{wtot}
	\end{minipage}
\end{figure}

The dependence of $W_c$ and $W_n$ on $E$ \eqref{wcnfinal0} is shown in Fig.~\ref{wcwn}. The solid curve corresponds to $W_c$ and the dotted curve corresponds to $W_n$. It is seen that there are  two peaks with the different  positions and heights, the distance between  peaks is  $\sim 2\,\mbox{MeV}$. This is a consequence of the Coulomb interaction, since in the absence of this interaction the peaks would coincide (recall that we did not take into account the small mass difference of $B^+$ and $B^0$). 
The width of each peak is approximately $17\,\mbox{MeV}$, and the  width of $W_{tot}=W_c+W_n$ is about $20\,\mbox{MeV}$, which is a consequence of the different positions of the peaks $W_c$ and~$W_n$. The energy dependence of $W_{tot}$ is shown in Fig.~\ref{wtot} by a dashed curve, and $W_{tot}$ averaged over the beam-energy spread of PEP-II is shown by a solid curve. The same figure shows experimental data from Ref.~\cite {Dong2020}, which are based on Ref.~\cite {BABAR2009} and take into account radiative corrections and radiative return. Assuming Gaussian distribution with $\Delta=4.6\,\mbox{MeV}$, averaging was carried out  according to the formula
\begin{equation}
\langle W(E)\rangle=\int_0^\infty W(E')\exp\left[-\dfrac{(E-E')^2}{2\Delta^2}\right]\dfrac{dE'}{\sqrt{2\pi}\,\Delta}\,.
\end{equation}
 One can see very good agreement between the predictions and the experimental data everywhere, except for the region above $35\,\mbox{MeV}$, which is most likely due to the close  threshold of $B^*\bar B$ and  $B\bar B^*$ pair  production.

\begin{figure}
	\begin{minipage}[t]{0.48\columnwidth}
		\centering
		\includegraphics[width=\linewidth]{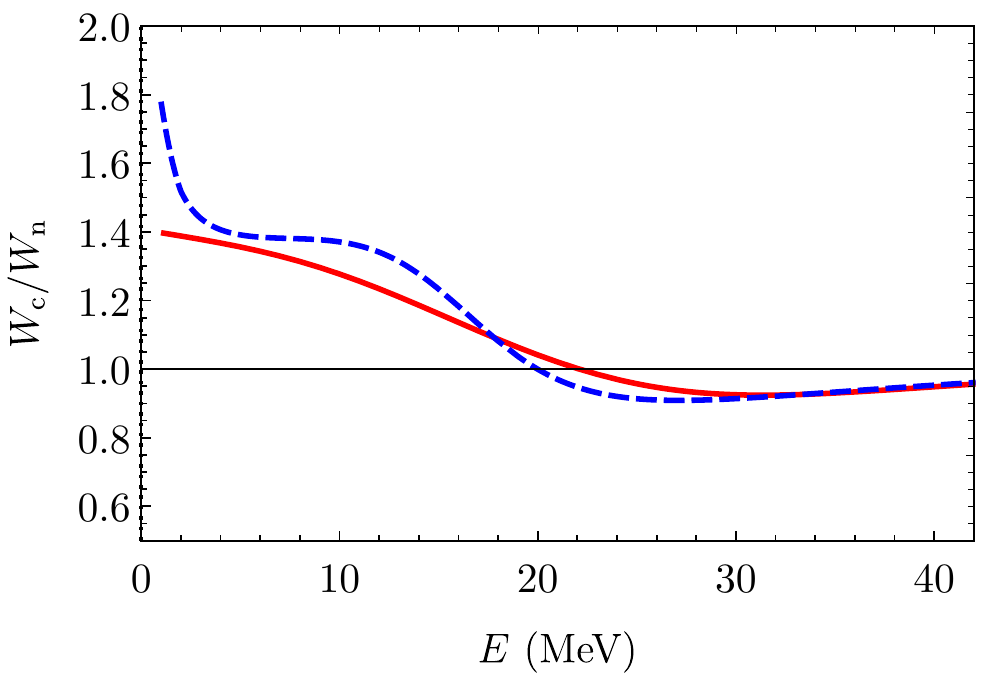}
		\caption{Energy dependence of the ratio $W_c/W_n $ (dashed curve) and $\langle W_c\rangle/\langle W_n\rangle $ (solid curve).}
		\label{wcwnratio}
	\end{minipage}
\hfill
	\begin{minipage}[t]{0.48\columnwidth}
		\centering
		\includegraphics[width=\linewidth]{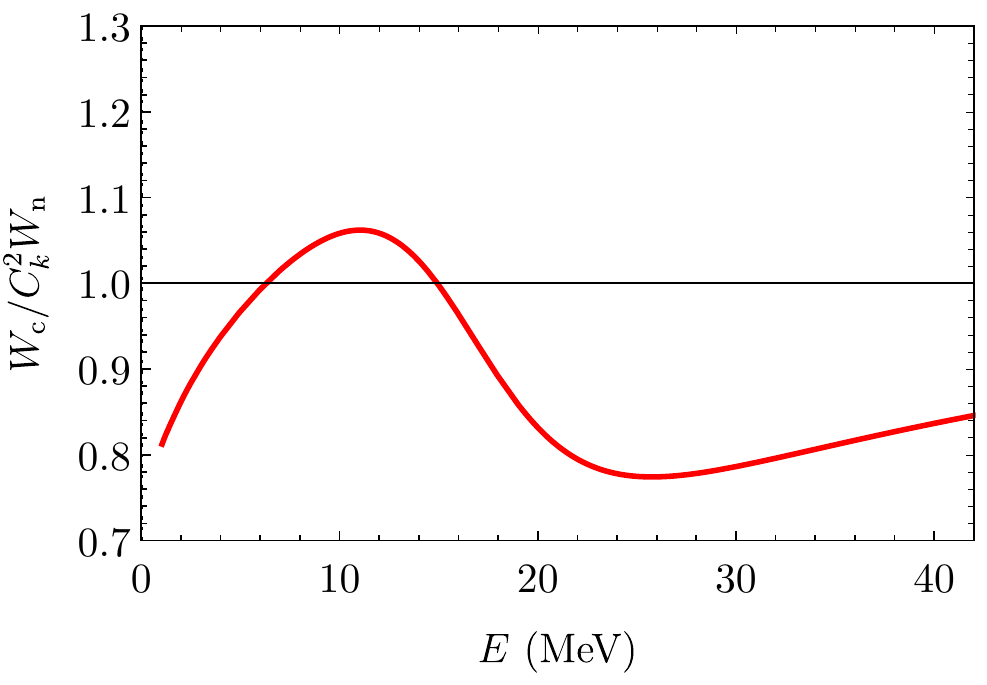}
		\caption{The energy dependence of $W_c/C^2_kW_n$.}
		\label{wcwnsommratio}
	\end{minipage}
\end{figure}

The energy dependence of the ratio $ W_c/W_n $ is shown in Fig.~\ref{wcwnratio}. The dotted curve corresponds to Eq.~\eqref{wcnfinal0} while the solid curve is the ratio of probabilities averaged over the beam-energy spread. It is seen that the ratio $W_c/W_n$ strongly depends on $E$, and near the maximum of $W_{tot}$ the function $W_c/W_n-1 $ passes through zero. The ratio $ W_c/W_n $ increases rapidly at lower energy and can reach 1.4 at  $E\sim 10\,\mbox {MeV}\,$.

The assumption that $W_c=C^2_k\,W_n$ is very often used to describe Coulomb effects. The energy dependence of the ratio $W_c/C^2_k\,W_n$ is shown in Fig.~\ref{wcwnsommratio}. It is seen that the hypothesis of factorization of the Coulomb effects does not work. Note that $W_n\propto k^3$ and  $W_c\to \mbox{const}\neq 0$ at  $E\to 0 $ so that $W_c/C^2_k\,W_n$ tends to a constant, but this constant is not equal to unity.
The violation of the factorization of Coulomb effects was first noted in \cite{MS2018} when describing the production of nucleon-antinucleon pairs in $e^+e^-$ annihilation near the threshold.

\section{Conclusion}
In our work, we have suggested a simple description of the decay probabilities $\Upsilon(4S)\to B^+B^-$ and $\Upsilon(4S)\to B^0\bar B^0$. Our results  are in good agreement with the available experimental data. We predict the existence of two peaks whose positions and heights differ due to Coulomb effects. Moreover, the ratio $W_c/W_n $ is a nontrivial function of energy, which increases rapidly as the energy decreases with respect to the peak position. It is also shown that the frequently used assumption of factorization of the Coulomb corrections is not in agreement  with the exact results.

\section*{Acknowledgments}
 We are grateful to A.E.~Bondar and R.V.~Mizuk for useful discussions.


\begin{thebibliography}{99}

\bibitem{CLEO2001} J.P. Alexander {\it et al.} [CLEO Collaboration], \href{https://doi.org/10.1103/PhysRevLett.86.2737}{Phys. Rev. Lett. {\bf 86}, 2737 (2001)}.

\bibitem{CLEO2002} S.B. Athar {\it et al.} [CLEO Collaboration], \href{https://doi.org/10.1103/PhysRevD.66.052003}{Phys. Rev. D {\bf 66}, 052003 (2002)}.

\bibitem{BABAR2002} B. Aubert {\it et al.} [BABAR Collaboraion], \href{https://doi.org/10.1103/PhysRevD.65.032001}{Phys. Rev. D {\bf 65}, 032001 (2002)}.

\bibitem{Belle2003} N.C. Hastings {\it et al.} [Belle Collaboraion], \href{https://doi.org/10.1103/PhysRevD.67.052004}{Phys. Rev. D {\bf 67}, 052004 (2003)}.

\bibitem{BABAR2004} B. Aubert {\it et al.} [BABAR Collaboraion], \href{https://doi.org/10.1103/PhysRevD.69.071101}{Phys. Rev. D {\bf 69}, 071101 (2004)}.

\bibitem{BABAR2005branching} B. Aubert {\it et al.} [BABAR Collaboraion], \href{https://doi.org/10.1103/PhysRevLett.95.042001}{Phys. Rev. Lett. {\bf 95}, 042001 (2005)}.

\bibitem{BABAR2005width} B. Aubert {\it et al.} [BABAR Collaboraion], \href{https://doi.org/10.1103/PhysRevD.72.032005}{Phys. Rev. D {\bf 72}, 032005 (2005)}.

\bibitem{BABAR2009} B. Aubert {\it et al.} [BABAR Collaboraion], \href{https://doi.org/10.1103/PhysRevLett.102.012001}{Phys. Rev. Lett. {\bf 102}, 012001 (2009)}.

\bibitem{Dong2020} Xiang-Kun Dong, Xiao-Hu Mo, Ping Wang, Chang-Zheng Yuan, \href{https://doi.org/10.1088/1674-1137/44/8/083001}{Chinese Phys. C {\bf 44}, 083001 (2020)}.

\bibitem{Belle2021} R. Mizuk {\it et al.} [Belle Collaboraion], \href{http://arxiv.org/abs/2104.08371}{arXiv:2104.08371 [hep-ex] (2021)}.

\bibitem{PDG2020} P.A. Zyla {\it et al.} (Particle Data Group), \href{https://doi.org/10.1093/ptep/ptaa104}{Prog. Theor. Exp. Phys. 2020, 083C01 (2020)}.

\bibitem{Marciano1990} D. Atwood and W.J. Marciano, \href{https://doi.org/10.1103/PhysRevD.41.1736}{Phys. Rev. D {\bf 41},  1736 (1990)}.

\bibitem{Lepage1990} G.P. Lepage, \href{https://doi.org/10.1103/PhysRevD.42.3251}{Phys. Rev. D {\bf 42}, 3251 (1990)}.

\bibitem{Eichten1990} N. Byers and E. Eichten, \href{https://doi.org/10.1103/PhysRevD.42.3885}{Phys. Rev. D {\bf 42},  3885 (1990)}.

\bibitem{Kaizer2003} R. Kaiser, A.V. Manohar, and T. Mehen, \href{https://doi.org/10.1103/PhysRevLett.90.142001}{Phys. Rev. Lett. {\bf 90}, 142001 (2003)}.

\bibitem{Voloshin2003} M.B.Voloshin, \href{https://doi.org/10.1142/S0217732303011538}{Mod. Phys. Lett. A {\bf 18}, 1783 (2003)}.

\bibitem{Voloshin2005} M.B.Voloshin, Yad. Fiz. {\bf 68}, 804 (2005) [\href{https://doi.org/10.1134/1.1935010}{Phys. At. Nucl. {\bf 68},  771 (2005)}].

\bibitem{DMS2014} V.F. Dmitriev, A.I. Milstein and S.G. Salnikov, Yad. Fiz. {\bf 77}, 1234 (2014) [\href{https://doi.org/10.1134/S1063778814080043}{Phys. At. Nucl. {\bf 77}, 1173 (2014)}].

\bibitem{DMS2016} V.F. Dmitriev, A.I. Milstein and S.G. Salnikov, \href{https://doi.org/10.1103/PhysRevD.93.034033}{Phys. Rev. D {\bf 93}, 034033 (2016)}.

\bibitem{MS2018} A.I. Milstein and S.G. Salnikov, \href{https://doi.org/10.1016/j.nuclphysa.2018.06.002}{Nucl. Phys. A {\bf 977}, 60  (2018)}.

\end{thebibliography}
 \end{document}